\documentclass[runningheads]{llncs}
\usepackage{graphicx}

\usepackage{hyperref}

\usepackage{xcolor}

\newif\ifshowComments
\showCommentstrue 

\ifshowComments
    \usepackage[draft,markup=bfit, deletedmarkup=sout, authormarkup=brackets]{changes}
\else
    \usepackage[final,markup=bfit, deletedmarkup=sout, authormarkup=brackets]{changes}
\fi
\definechangesauthor[name={Kassandra},color=green]{KF}
\definechangesauthor[name={Matej}, color=orange]{MH}
\definechangesauthor[name={Kika}, color=cyan]{KM}
\definechangesauthor[name={Sabina}, color=magenta]{SS}
\ifshowComments

\else

\fi

\usepackage{draftwatermark}

\begin{document}
\SetWatermarkAngle{0}
\SetWatermarkColor{black}
\SetWatermarkLightness{0.5}
\SetWatermarkFontSize{10pt}
\SetWatermarkVerCenter{25pt}
\SetWatermarkText{\parbox{30cm}{%
\centering This is the authors' final version of the manuscript published as\\
\centering Friebe, K., Malinovská, K., Samporová, K., Hoffmann, M. (2022). Gaze Cueing and the Role of Presence in Human-Robot Interaction.\\
\centering In: Social Robotics. ICSR 2022. Lecture Notes in Computer Science, vol 13817. Springer, Cham, pp. 402-414. \\
\centering \url{https://doi.org/10.1007/978-3-031-24667-8_36}
}}

\title{Gaze Cueing and the Role of Presence in Human-Robot Interaction}

\titlerunning{Gaze Cueing and the Role of Presence in Human-Robot Interaction}

\author{Kassandra Friebe\inst{1,2}\orcidID{0000-0002-6208-3310} \and
Sabína Samporová\inst{2} \and
Kristína Malinovská\inst{2}\orcidID{0000-0001-7638-028X} \and
Matej Hoffmann\inst{3}\orcidID{0000-0001-8137-3412}
}
\authorrunning{K. Friebe et al.}

\institute{Department of Cognitive Science, Central European University, Vienna, Austria\\ \and
Faculty of Mathematics, Physics and Informatics, Comenius University,\\Mlynská dolina,  842 48 Bratislava, Slovakia\\ \and
Faculty of Electrical Engineering, Czech Technical University in Prague,\\Technická 1902/2, 166 27 Praha 6 - Dejvice, Czechia\\
\email{kassandra.friebe@web.de}\\
}
\maketitle              

\begin{abstract}
Gaze cueing is a fundamental part of social interactions, and broadly studied using Posner task based gaze cueing paradigms. While studies using human stimuli consistently yield a gaze cueing effect, results from studies using robotic stimuli are inconsistent. Typically, these studies use virtual agents or pictures of robots. As previous research has pointed to the significance of physical presence in human-robot interaction, it is of fundamental importance to understand its yet unexplored role in interactions with gaze cues. This paper investigates whether the physical presence of the iCub humanoid robot affects the strength of the gaze cueing effect in human-robot interaction. We exposed 42 participants to a gaze cueing task. We asked participants to react as quickly and accurately as possible to the appearance of a target stimulus that was either congruently or incongruently cued by the gaze of a copresent iCub robot or a virtual version of the same robot. Analysis of the reaction time measurements showed that participants were consistently affected by their robot interaction partner's gaze, independently on the way the robot was presented. Additional analyses of participants' ratings of the robot's anthropomorphism, animacy and likeability further add to the impression that presence does not play a significant role in simple gaze based interactions. Together our findings open up interesting discussions about the possibility to generalize results from studies using virtual agents to real life interactions with copresent robots.

\end{abstract}
%
%
%
\section{Introduction}
Human social interactions are based on a complex exchange of a variety of signals, including facial expressions, gaze, gestures, and posture. Gaze plays a special part as it serves to perceive objects or other humans and at the same time it communicates to others where one focuses the attention \cite{risko2016}. 
For instance, our gaze helps us indicate
social interest \cite{stass1967eye}, understand other people's mental and emotional state \cite{baron1997there}, and see what they are attending to \cite{frischen2007}. This process of using someone
else’s eye movement as information of what they are attending to and shifting one’s own
attention accordingly is called gaze cueing and is discussed as a prerequisite for joint
attention \cite{emery1997gaze}, the case in which both persons visually attend the
same object. 
\\

Social gaze has been mostly studied in human and non-human primates, and with the emergence of social robots has become introduced into robotic systems
\cite{breazeal1999,kozima1998towards} and has become a growing branch of research ever since. A great amount of research on gaze cueing in human-robot interaction (HRI) uses virtual agents or pictures of robots instead of physically copresent robots.
From other work on human-robot interaction, we know however, that the physical presence of a robot fundamentally shapes the way we perceive and interact with it \cite{bainbridge2008effect,li2015benefit,wainer2006role}. 
What we do not know, however, is how the physical presence of an agent and gaze cueing relate. This paper thus explores whether the physical presence of an agent affects the strength of the gaze cueing effect in human-robot interaction.
\\
\section{Related Work}
The common paradigm used to study gaze cueing is a variation of the Posner paradigm \cite{posner1980orienting}, in which participants are asked to localize a target stimulus while that stimulus is consistently and inconsistently cued by a facial stimulus. In several studies using schematic and human faces as stimuli, a gaze cueing effect (GCE) has been observed, evidenced by faster reaction times in responding to congruent (gazed-at) target stimuli compared to incongruent (non-gazed-at) target stimuli \cite{friesen1998eyes,driver1999gaze}. 

In contrast to these common findings, studies using robotic stimuli show mixed results. While Admoni and colleagues \cite{admoni2011robot} find a gaze cueing effect with images of human, schematic, and no GCE for the robot faces in one study, 
Wiese and colleagues \cite{wiese2014using} report a more pronounced GCE for robotic face stimuli compared to human face stimuli in a study with individuals with autism spectrum disorder (ASD).
Notably, most studies on gaze cueing in HRI used only static images of gazing robots. For instance, in a study investigating the effect of human-likeness of a robot on the strength of the gaze cueing effect, Martini et al. used morphed images of human and robot faces as stimuli~\cite{martini2015}. To date, however, it is still unclear to what extent these results can be generalized to copresent robots in real human-robot interaction.

Physical presence appears to be an important factor shaping our perception of and interaction with robots \cite{li2015benefit}. For example, in an early study conducted by Lee et al. \cite{lee2006physically}, in which participants were either introduced to a physical Aibo robot dog, or its virtual equivalent, a positive effect of embodiment on ratings of the interaction with the robot and the robot’s social presence was found. These results indicate the importance of physical embodiment in HRI, even if it is not necessary to complete the interaction successfully.

Wiese and colleagues employed a gaze cueing paradigm using a copresent humanoid robot \cite{wiese2018}. Participants were instructed to indicate the appearance of a light stimulus on their left or right side, which was cued by a Meka robot's gaze shift. 
Even though the researchers informed the participants that the robot's gaze is uninformative of the appearance of the light stimulus, participants seemed to follow the robot's gaze, as a congruency effect could still be found. Using a similar setup, Kompatsiari et al. \cite{kompatsiari2018role} were able to replicate these results with an iCub robot when the robot established mutual gaze with the observer before turning to the target stimuli, as in the previously reported experiment \cite{wiese2018}, but not when no mutual gaze was established.


Further controlled investigations will be needed, in which participants are faced with scenarios and robots that are constant except for the way they are presented, to better grasp the role of presence on gaze cueing effects in HRI. To our knowledge, there is only one systematic study conducted to date that explores the relation of embodiment, presence and facial cueing. Mollahosseini et al.~\cite{mollahosseini2018} investigated the effect of robot embodiment and presence on interaction tasks that involved typical measures of communication, and found that these factors affect recognition of facial expressions, and especially eye gaze. While further analysis of the data revealed a significant effect of embodiment, unlike other studies examining social interaction in HRI, no main effect of presence was found. This could be due to the nature of the task, which involved only still representations of the gaze rather than actual movements, or to the fact that it generally did not rely on social attributions that could relate to the presence of the robot, but on purely geometric cues.

\section{Methods and Material}
\begin{figure}
\includegraphics[width=\textwidth]{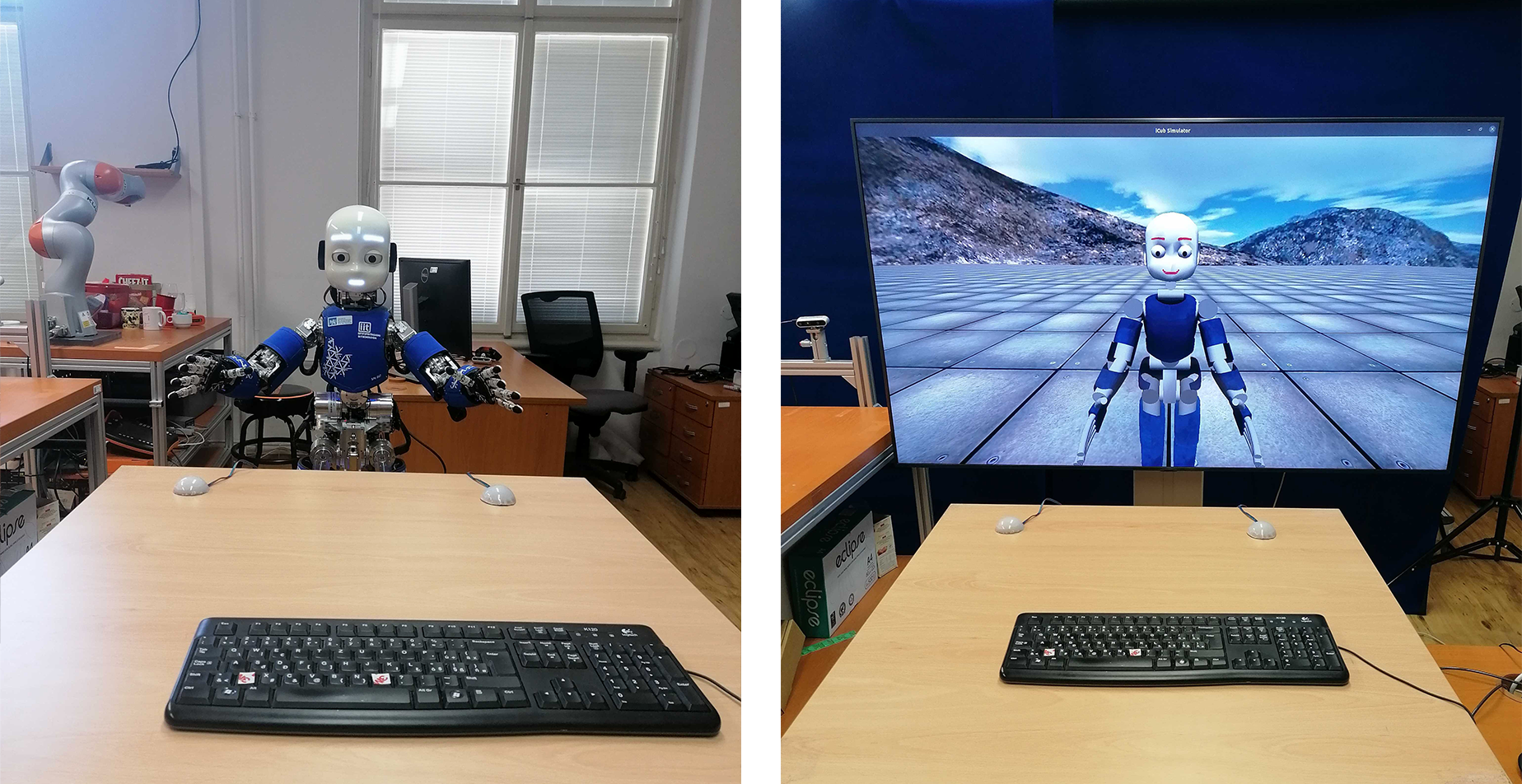}
\caption{Copresent (left) and virtual iCub (right) used in the experiment.} 
\label{figure2}
\end{figure}

We chose a mixed experimental design with two independent variables: (1) type of robot presence (between-subject) with two levels: a physical robot and a virtual version of the same robot as depicted in Fig.~\ref{figure2} (2) gaze cue congruency (within-subject) with two levels: congruent, and incongruent as depicted in Fig.~\ref{figure3}. Participants were randomly assigned to one type of robot and observed it gazeing at one of two light stimuli situated on either side of the table. The robot shifted its gaze in 80 trials, 40 congruent trials in which the robot looked at the lamp that was to change color and 40 incongruent trials in which the robot looked at the opposite lamp (see Fig.~\ref{figure3}). In half of the conditions the robot looked at the left side and in the other half at the right side. All conditions were pseudo-randomly shuffled. Most of the features detailed in this section are also illustrated in this video \url{https://youtu.be/n_rU9XNE-bI}.

\subsection{Participants}
A total of 42 participants were recruited via posters in the university building, Facebook advertisement, and email. 
Three participants were excluded due to technical problems with the setup. 

The final sample consisted of 42 participants (16 females; mean age = 29.98). All participants provided written informed consent in line with the ethical approval of the study granted by the Committee for Research Ethics at the Czech Technical University in Prague.
When asked about their experience with robots on a scale from 1 (very poor) to 5 (very good), the mean score was 2.33 (SD = 1.2) and only one participant answered 5. Data was stored and analysed anonymously. Testing time was about fifteen minutes.

\subsection{Measures}
Similar to previous studies on gaze cueing \cite{wiese2018,driver1999gaze},
the influence of gaze cueing on participants’ gaze following behavior was determined
by measures of mean correct reaction time. A response was considered incorrect if
it was made with the wrong key press, and considered correct if the correct key was
pressed. Responses that were given in a response time that was more than 2.5 standard
deviations away from the individual mean response time of a participant were excluded
from further analyses. To further check for possible effects of a robot’s physical presence
on gaze following behavior, average correct response times (RTs) were calculated for
each participant and each experimental condition.

\subsection{Experimental Set-Up and Procedure}
The present experiment was designed to examine both self-reported and behavioral effects of a robot's presence and gaze cues in a human-robot interaction task. During the experiment, participants were instructed to indicate the appearance of a target stimulus that was either congruent or incongruent with the position being gazed at by an iCub robot that was either physically present in the same room with the participants (copresence condition) or presented via a monitor (virtual agent condition) by pressing a corresponding key on a keyboard. After completion of the task, participants were asked to indicate the way they perceived the robot by completing the Czech translation of the three subscales ``Anthropomorphism'', ``Animacy'' and ``Likeability'' of the Godspeed series \cite{bartneck2009measurement}\footnote{
\url{https://www.bartneck.de/2008/03/11/the-godspeed-questionnaire-series/}}. 
The experiment was conducted in April 2022 in the laboratories of the Department of Cybernetics of the Czech Technical University in Prague. 

At the beginning of the experiment, participants received written and oral instructions and gave informed consent. They were informed that the task was to respond as fast as possible to the color change of one of two 
light stimuli. Responses had to be given by pressing the appropriate key on a keyboard. Participants were also informed that iCub might move randomly during the time of the experiment. Reaction times were measured as a dependent variable. After receiving the instructions participants had the opportunity to ask questions about the task.

\begin{figure}[ht]
\includegraphics[width=6cm]{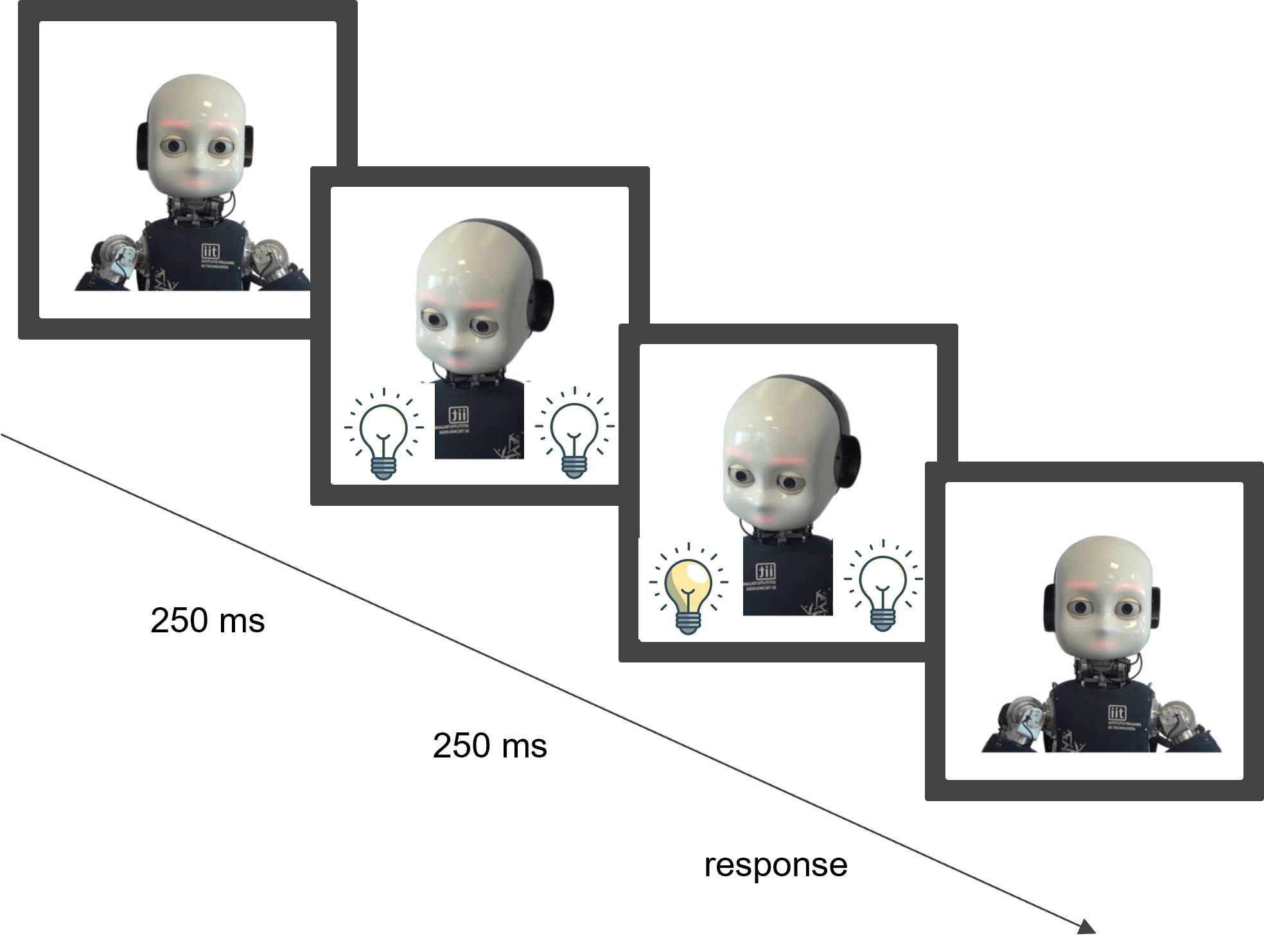}
\includegraphics[width=6cm]{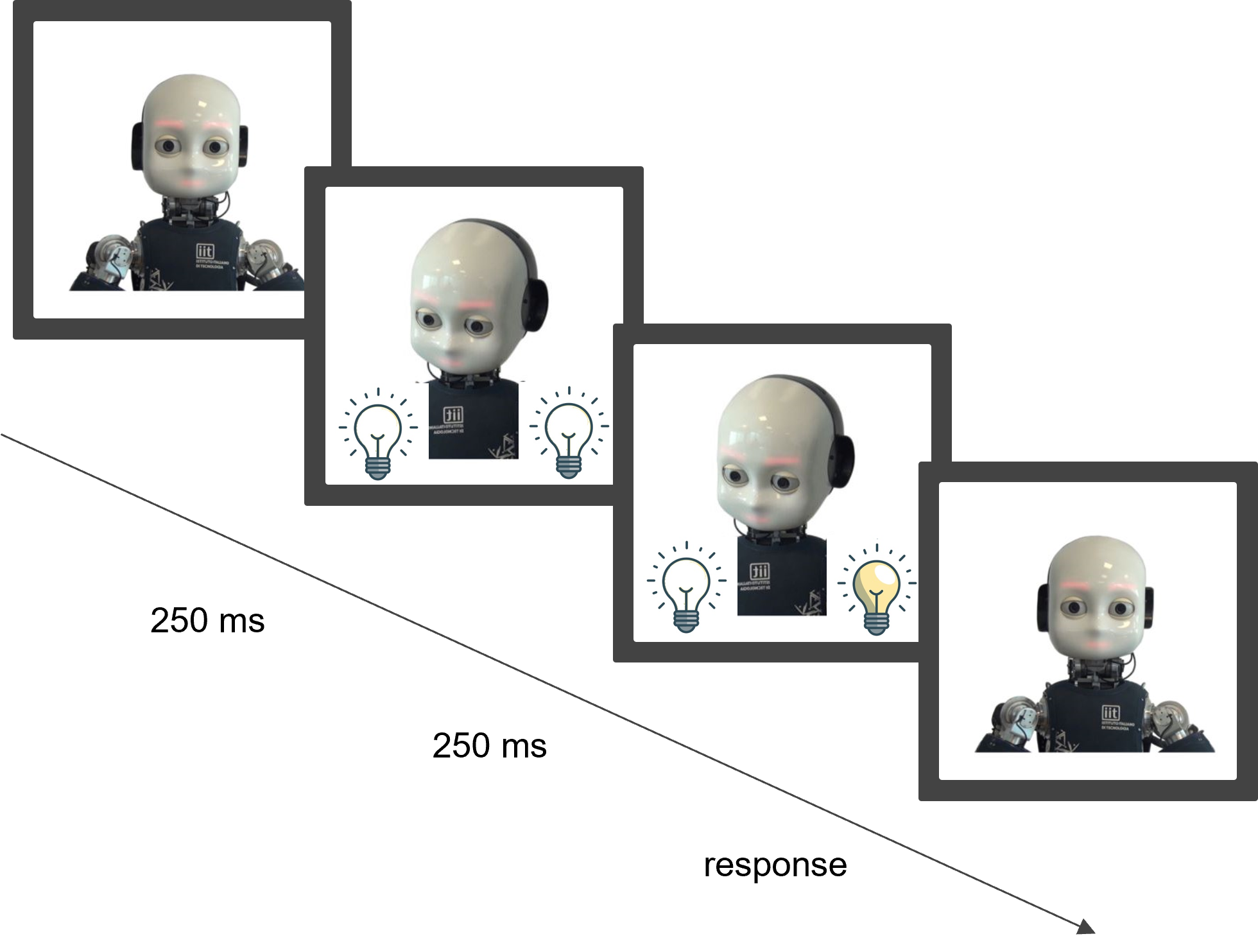}
\caption{Example trial sequence with (left) congruent condition and (right) incongruent condition for the physically presence condition. Target stimuli are represented as schematic depiction (light bulb vs. self-made light stimuli in the actual experiment).} 
\label{figure3}
\end{figure}

Each trial began with iCub making eye contact by looking straight ahead in the direction of the observer. After 250 ms, iCub shifted his gaze either toward the lamp that was on his left side or toward the lamp that was on his right side. Subsequently, after 200 ms, one of the two lamps changed color, either on the corresponding side of the gaze cue or on the non-corresponding side of the gaze cue. When the target stimulus was presented, participants responded as quickly and as accurately as possible to the position of the target stimulus by pressing the "x" or "m" key on a standard keyboard. The target stimulus remained unchanged until a response was made or a time-out criterion (5000 ms) was reached. Then the light was turned off again and iCub looked straight ahead again to signal readiness to begin the next trial. Figure~\ref{figure3} shows an exemplary trial sequence.

\subsection{Hardware and Software}
The iCub \cite{metta2010icub} is a small humanoid robot that resembles a 4-year old child. It is one meter tall and has 53 Degrees of Freedom (DoF). Most relevant to gaze, it has 3 DoF in the neck and 3 coupled DoF for the two eyes (tilt, version, and vergence) in an anthropomorphic arrangement. 
For the virtual iCub we used the freely available simulator \cite{tikhanoff2008open}.
The iCub gaze controller \cite{roncone2016cartesian} is used to command where the robot should look.
Our custom made C++ program that controls the movement of the robot using the YARP middleware and records the reaction times in the experiment is publicly available\footnote{\url{github.com/Sabka/icub-hri-cuing}}.
For our experiment we designed custom lamps which consisted of mate covered red led lamps controlled via Arduino Nano.

\subsection{Analysis}
Statistical analyses were conducted using R (version 4.1.2). 
The average correct RTs
for congruent and incongruent trials were compared for each robot presence condition
individually to check for consistency with previous studies regarding the strength of
the gaze cueing effect. To test for the effect of cue-target congruence, for the virtual
robot condition a t-test was calculated comparing the mean reactions times of congruent trials and incongruent trials. For the copresent robot condition a Wilcoxon test
was calculated instead, to account for non-normally distributed data. A gaze cueing effect is evidenced by significant differences in reaction times of congruent and incongruent trials. 

Moreover, a mixed model analysis of variance (ANOVA) with
a between-subject factor of robot presence (2: copresence vs. virtual presence), and
within-subject factors of cue-target congruency (2: congruent vs. incongruent) was
calculated. This analysis was used to assess the individual effect of cue-target congruency and the effect of presence specificity of gaze cueing. Robot presence specific
gaze-cueing effects would be evidenced by a significant interaction between presence
condition and cue-target congruency (over and above a main effect of congruency). By
contrast, presence nonspecific gaze cueing would manifest in terms of a main effect of
cue-target congruency (not accompanied by a presence~$\times$ congruency interaction), with
equal facilitation for all robot presence conditions of interest. 

\section{Results}
\subsection{The Effect of Presence and Gaze}
Mean reaction times were subjected to a two-way analysis of variance with two levels of robot presence (copresent robot, virtual agent) and two levels of cue-target congruence (congruent, incongruent) to test the effect of robot physical presence and gaze cueing behavior on participants' reaction times in a localization task (see Table~\ref{table1}). It is important to note that the data was not normally distributed in each group. An ANOVA was conducted either way, as F-Tests have been reported to be relatively robust to violations of normality when homogeneity of variance is given \cite{blanca2017non}.

\begin{table}[ht]
\centering
\caption{Gaze Congruence x Robot Presence Analysis of Variance}
\begin{tabular}{lllllllll}
\hline
Source          & Df & F     & $\mu$2   & p          \\\hline
Robot Presence  & 1  & 0.64  & 0.02 & 0.43       \\
Gaze Congruence & 1  & 18.71 & 0.32 & .001***    \\
Presence x Gaze & 1  & 2.22  & 0.05 & 0.14       \\
Error           & 40 &       &      &           \\\hline
\end{tabular}
\\\vspace{0.2em}\emph{Note. *** p$<$.001, ** p$<$.01, * p$<$.05}
\label{table1}
\end{table}
\noindent

The main effect of gaze congruency yielded an F-ratio of F(1, 40) = 18.71, p~$<$~.001, indicating that mean reaction times differed significantly between congruent and incongruent trials, with faster reaction times on congruent cue-target trials (M = 288 ms, SD = 37.2 ms) than on incongruent cue-target trials (M = 298 ms, SD = 43.2 ms). 

\begin{figure}
\includegraphics[width=0.5\textwidth]{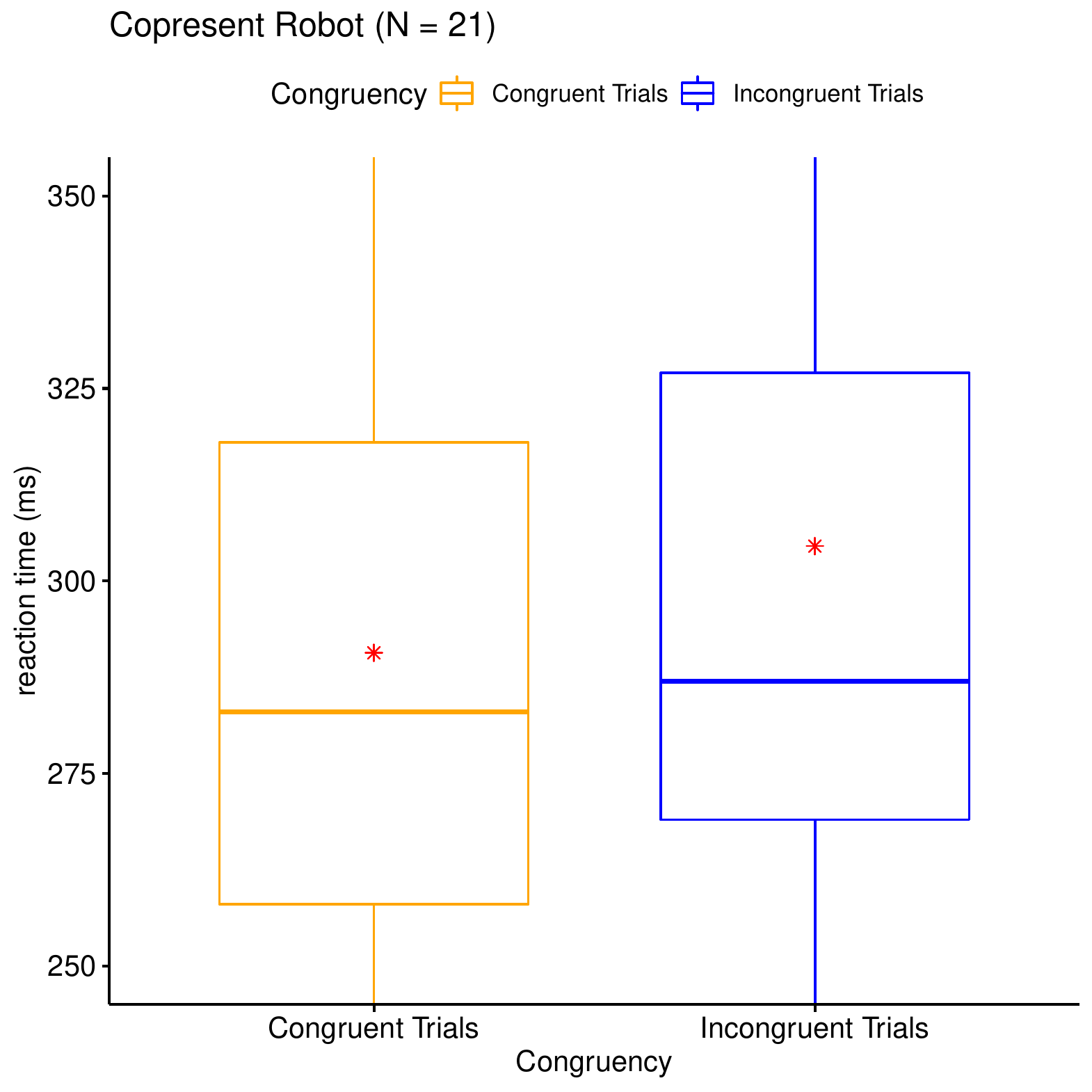}
\includegraphics[width=0.5\textwidth]{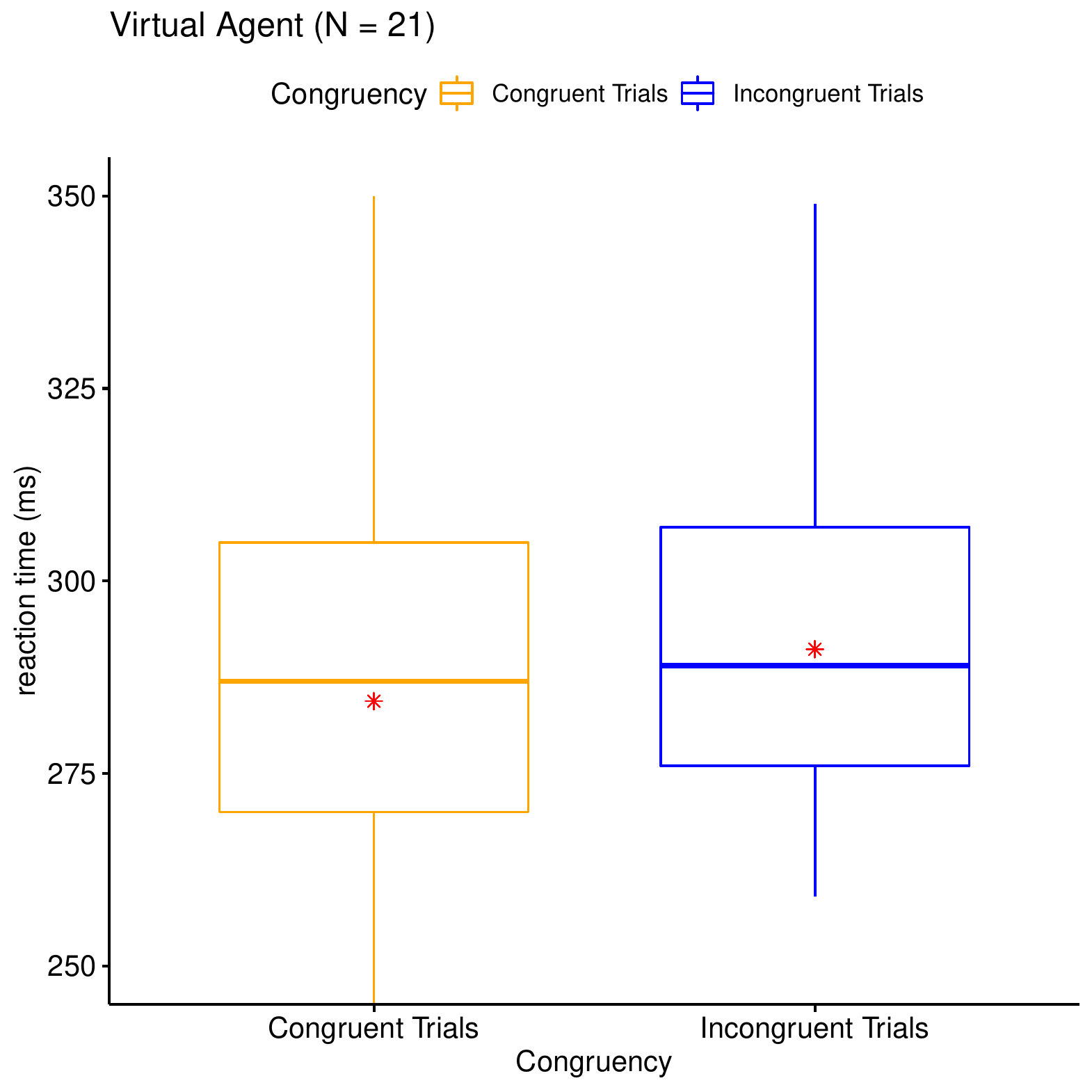}
\caption{Mean reaction times by robot presence group. Values are reported in milliseconds. Group means are illustrated as red stars.} \label{figure4}
\end{figure}

To illustrate the size of the GCE by robot presence group, individual analyses were performed for both robot conditions as displayed in Figure~\ref{figure4}. For the comparison of mean reaction times in congruent and incongruent trials in the copresent robot group a Wilcoxon signed-rank test was calculated to account for non-normally distributed data as indicated by a significant
Shapiro-Wilk test (W~=~.9, p~=~.009). On average, participants in the copresent robot condition responded faster on congruent trials (M~=~291 ms) than on incongruent trials (M~=~305 ms). Results of the Wilcoxon signed-rank test showed that this difference was statistically significant (p~=~.001), with a large effect size, r = 0.7. An additional t-test conducted on the mean reaction times of participants in the virtual agent group revealed a significant difference between congruent and incongruent trials (t (20) = -2, p = .03) with shorter reaction times for congruent (M = 284 ms) than incongruent trials (M = 291 ms). The effect size was at a moderate level, r = 0.5.

While participants in the virtual agent group (M = 287 ms) on average reacted faster than participants in the copresent group (M = 298 ms), the main effect of robot presence on participants' mean reaction times was non-significant F(1, 40) = 0.64, p $>$ .05. Yet, it is worth noting that the variances of the reaction times of the two robot conditions were significantly different, p =.012. Moreover, no significant interaction effect of gaze congruency and presence could be found, F(1, 40) = 2.22, p $>$ .05.

\subsection{Analysis of Godspeed Indices}
\begin{figure}[ht]
\centering
\includegraphics[width=0.65\textwidth]{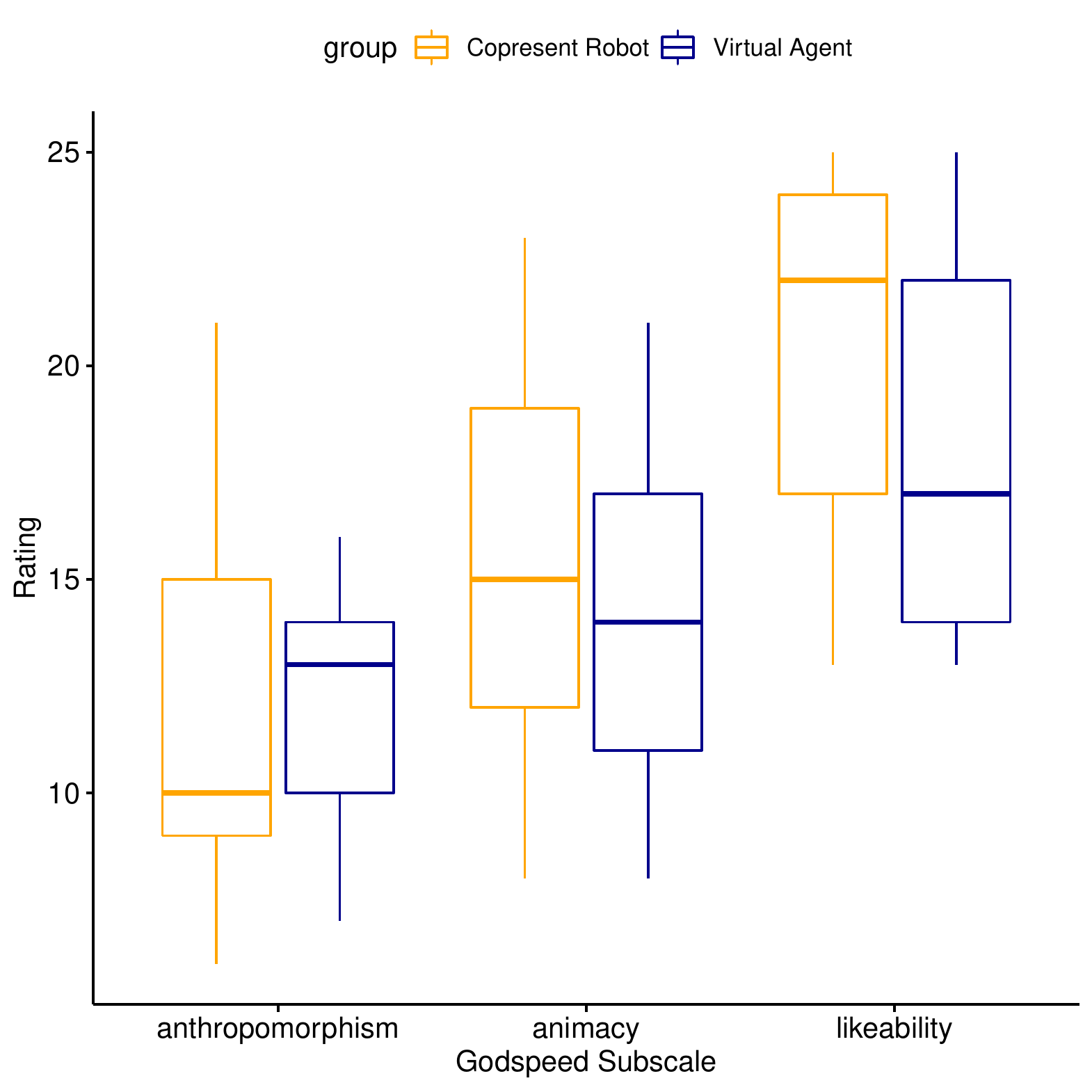}
\caption{Ratings of the Godspeed Indices by Group. Scores are summed for each individual subscale. The Anthropomorphism and Likeability scale consist
of 5 sub-questions on a 5-point Likert scale with a maximum score of 25 points, and the Animacy scale consists of 6 sub-questions on a 5-point Likert scale with a maximum
score of 30 points.} 
\label{figure5}
\end{figure}

To further test how robot presence influences the way participants perceive the robot, responses of the participants to the Godspeed subscales Anthropomorphism, Animacy and Likeability were taken into account. A standard t-test was used to examine the influence of robot presence on animacy ratings. Significant results in a Shapiro-Wilk test with mean ratings of anthropomorphism and likability as outcome variables indicated non-normally distributed data, so an unpaired two-samples Wilcoxon test was computed to test the influence of robot presence on perceived anthropomorphism and likability. Mean ratings for all three Godspeed subscales by robot presence condition can be found in Figure~\ref{figure5}.

\subsubsection{Anthropomorphism}
For ratings of the robot’s perceived anthropomorphism, participants in the virtual agent condition assessed the robot’s anthropomorphism slightly higher (M = 12.29, SD = 2.83) than people in the copresent robot condition (M = 12.05, SD = 4.03). Results of the independent samples Wilcoxon test, however, were not significant; W = 208.5, p = .77, r = -0.05. A linear model including sex as an additional predictor was tested due to the significant correlation of anthropomorphism ratings and sex. The model revealed no significant difference between presence groups, whereas ratings between sexes significantly differed (p $<$ .05), with females (M = 13.56, SD = 3.44) rating the robot as more anthropomorphic on average than males (M = 11.31, SD = 3.21). 

\subsubsection{Animacy}
On average, participants in the copresent robot condition rated the robot’s perceived animacy higher (M = 15.19, SD= 4.50) than participants in the virtual agent group (M = 13.95, SD = 3.67). This difference was not significant; t(38.43) = 0.98, p = .33. 

\subsubsection{Likeability}
Participants in the copresent robot condition assessed the robot’s likeability slightly higher (M = 20.19, SD = 4.25) than people in the copresent robot condition (M = 18.10, SD = 3.94). Results of the independent samples Wilcoxon test indicate that this difference was not significant; W = 289, p = .09, r = 0.26. A linear model including sex as an additional predictor was tested due to the significant correlation of likeability ratings and sex. The model revealed no significant difference between presence groups, whereas ratings between sexes significantly differed (p $<$ .01). On average, females (M = 21.23, SD = 3.41) rated the robot as more likeable than males (M = 17.84, SD = 4.13).

\section{Discussion}
Our results are consistent with previous research on gaze following with copresent robots \cite{wiese2018}: Participants consistently exhibited gaze-following behavior, as evidenced by slower reaction times on trials in which the robot cued the wrong target compared to trials in which the cued location and target location matched. Hence, our results replicate the well-known finding that participants locate a target that is congruent with the cued direction more quickly than a target that is incongruent with the cued direction \cite{friesen1998eyes,driver1999gaze}. In particular, our results are in line with findings by Wiese \cite{wiese2018} and by Kompatsiari \cite{kompatsiari2018role} showing that a gaze cueing effect can be found when the target stimulus is predicted by the gaze of an embodied robot. This finding is particularly valuable given the ongoing replication crisis in psychological research, which has highlighted the problem of replicating the results of many scientific studies \cite{open2015estimating} and allows for the generalization of the gaze cueing effect across different robotic platforms. Crucially, with an effect of 14 ms, the gaze cueing effect in our study is smaller than the 25 ms found by \cite{wiese2018} but in line with findings of other studies using more controlled settings \cite{wiese2012see,wykowska2014beliefs}. Differences in the extent of the effect might be explained by the design of the robot---possibly, the Meka robot used in \cite{wiese2018} offers more social cues or other affordances than the iCub robot used in the present study or other robotic platforms. Further studies will be needed to better understand the relationship of robot design and gaze following behavior. Moreover, our results show that even the virtual version of our robot consistently triggered a gaze cueing effect as indicated by slower reaction times in incongruent compared to congruent trials, showing that gaze following in HRI is not limited to physically present versions of embodied robots.

Importantly, however, the same stimuli did not elicit varying degrees of gaze-cueing when comparing the different ways the robot was presented to the participants, as evidenced by a non-significant interaction of robot presence and cue-target congruency. Across conditions participants consistently followed the gaze, independently of whether they were confronted with a copresent iCub or a virtual iCub. As the novelty of this study lies in comparing the way the different ways of presenting the robot influence simple social attention mechanisms, as evidenced by the gaze cueing effect, there exists hardly any literature indicating similar or contradictory results. However, these results add to findings of Mollahosseini \cite{mollahosseini2018} who were comparing the effect of embodiment and presence of four different types of agents on similar outcome variables. Results showed that embodiment but not physical presence was the factor that accounts for the significant difference in the participants’ response, as indicated by no significant difference in results when presenting the participants with a copresent robot compared to a telepresent robot. However, the outcomes differed significantly for the comparisons of virtual agents and both forms of embodied robots. In contrast, our study found no significant difference in gaze following behavior between the presence conditions, as participants' reaction times were not significantly different when interacting with a virtual agent compared to a copresent robot.

There are multiple possible reasons why physical presence might not additionally influence the gaze cueing effect. One could be that our results are based on the fact that robots in both conditions moved in a similar, human-like manner. Previous research has shown that (natural) movement is linked to mind attributions---famously for example in the Heider and Simmel illusion~\cite{heider1944experimental}, in which participants attribute mental states to three moving geometrical figures (two triangles that seem to ``hunt'' a circle). Studies examining how mental state attributions alter gaze following behavior in trials with photographs of robots have shown that this manipulation led to the occurrence of a gaze cueing effect that was not otherwise present \cite{wykowska2014beliefs}. Differences between our results and those of studies using only photos of agents or robots \cite{admoni2011robot} or non-moving agents, as in the research of Mollahosseini \cite{mollahosseini2018}, might be due to the association of motion and mind attributions leading to typical social interaction phenomena, such as gaze cueing. 

Moreover, in contrast to previous findings \cite{li2015benefit}, the copresent robot did not generate more positive ratings than the virtual agent, as indicated by participants' ratings of the robot's anthropomorphism, animacy, and likability. Similar results were reported in a study \cite{kiesler2008anthropomorphic}, in which the ratings of four types of agents that differed in terms of their embodiment and physical presence were compared after a simple conversational interaction. Notably, neither the interaction reported by \cite{kiesler2008anthropomorphic} nor the interaction reported in our study involved physical touch or a particular focus on spatial relations. The differences between the results of this and other studies could be explained by the advantages of the physical presence of a robot in more complex interaction scenarios.

\section{Conclusion}
Social gaze is a fundamental part of human interactions and becomes more relevant in the scope of social human-robot interaction. Gaze cueing, the event in which we observe our interaction partner's gaze and shift our own attention accordingly, is broadly studied using images or virtual versions of robots. As previous research has pointed to the broad range of effects physical presence has in human-robot interaction, the question arises whether it affects the strength of the gaze cueing effect and hence, whether results from studies using images or virtual agents can be generalized to copresent robots. We designed a study to investigate the relationship of physical presence and social gaze by adapting a gaze cueing paradigm with two types of agents (1) a copresent robot and (2) a virtual version of the same robot. The results of our study indicate that gaze cueing is a stable effect in basal human-robot interaction across different robot presence conditions. Thereby, we add to the understanding of possibilities to generalize results from studies using virtual agents and pictorial stimuli to real life human-robot interaction.

\subsubsection*{Acknowledgements}
This research was supported by the by the Czech Science Foundation (GA\v{C}R), project 20-24186X. The authors from Comenius University were additionally supported by the Slovak Research and Development Agency, project APVV-21-0105 and by Slovak Society for Cognitive Science. We would like to dearly thank Jakub Rozlivek and Lukáš Rustler for their help in programming the iCub and Adam Rojík for organizing the recruitment of participants.
%
%
%
\bibliographystyle{splncs04}
\bibliography{mybibliography}

\begin{thebibliography}{10}
\providecommand{\url}[1]{\texttt{#1}}
\providecommand{\urlprefix}{URL }
\providecommand{\doi}[1]{https://doi.org/#1}

\bibitem{admoni2011robot}
Admoni, H., Bank, C., Tan, J., Toneva, M., Scassellati, B.: Robot gaze does not
  reflexively cue human attention. In: Proceedings of the Annual Meeting of the
  Cognitive Science Society. vol.~33 (2011)

\bibitem{bainbridge2008effect}
Bainbridge, W.A., Hart, J., Kim, E.S., Scassellati, B.: The effect of presence
  on human-robot interaction. In: RO-MAN 2008-The 17th IEEE International
  Symposium on Robot and Human Interactive Communication. pp. 701--706. IEEE
  (2008)

\bibitem{baron1997there}
Baron-Cohen, S., Wheelwright, S., Jolliffe, T.: Is there a" language of the
  eyes"? evidence from normal adults, and adults with autism or asperger
  syndrome. Visual cognition  \textbf{4}(3),  311--331 (1997)

\bibitem{bartneck2009measurement}
Bartneck, C., Kuli{\'c}, D., Croft, E., Zoghbi, S.: Measurement instruments for
  the anthropomorphism, animacy, likeability, perceived intelligence, and
  perceived safety of robots. International journal of social robotics
  \textbf{1}(1),  71--81 (2009)

\bibitem{blanca2017non}
Blanca~Mena, M.J., Alarc{\'o}n~Postigo, R., Arnau~Gras, J., Bono~Cabr{\'e}, R.,
  Bendayan, R., et~al.: Non-normal data: Is anova still a valid option?
  Psicothema  (2017)

\bibitem{breazeal1999}
Breazeal, C., Scassellati, B.: How to build robots that make friends and
  influence people. In: Proceedings 1999 IEEE/RSJ International Conference on
  Intelligent Robots and Systems. Human and environment friendly robots with
  high intelligence and emotional quotients. vol.~2, pp. 858--863. IEEE (1999)

\bibitem{driver1999gaze}
Driver~IV, J., Davis, G., Ricciardelli, P., Kidd, P., Maxwell, E., Baron-Cohen,
  S.: Gaze perception triggers reflexive visuospatial orienting. Visual
  cognition  \textbf{6}(5),  509--540 (1999)

\bibitem{emery1997gaze}
Emery, N.J., Lorincz, E.N., Perrett, D.I., Oram, M.W., Baker, C.I.: Gaze
  following and joint attention in rhesus monkeys (macaca mulatta). Journal of
  comparative psychology  \textbf{111}(3), ~286 (1997)

\bibitem{friesen1998eyes}
Friesen, C.K., Kingstone, A.: The eyes have it! reflexive orienting is
  triggered by nonpredictive gaze. Psychonomic bulletin \& review
  \textbf{5}(3),  490--495 (1998)

\bibitem{frischen2007}
Frischen, A., Bayliss, A.P., Tipper, S.P.: Gaze cueing of attention: visual
  attention, social cognition, and individual differences. Psychological
  bulletin  \textbf{133}(4), ~694 (2007)

\bibitem{heider1944experimental}
Heider, F., Simmel, M.: An experimental study of apparent behavior. The
  American journal of psychology  \textbf{57}(2),  243--259 (1944)

\bibitem{kiesler2008anthropomorphic}
Kiesler, S., Powers, A., Fussell, S.R., Torrey, C.: Anthropomorphic
  interactions with a robot and robot--like agent. Social Cognition
  \textbf{26}(2),  169--181 (2008)

\bibitem{kompatsiari2018role}
Kompatsiari, K., Ciardo, F., Tikhanoff, V., Metta, G., Wykowska, A.: On the
  role of eye contact in gaze cueing. Scientific reports  \textbf{8}(1),  1--10
  (2018)

\bibitem{kozima1998towards}
Kozima, H., Ito, A.: Towards language acquisition by an attention-sharing
  robot. In: New methods in language processing and computational natural
  language learning (1998)

\bibitem{lee2006physically}
Lee, K.M., Jung, Y., Kim, J., Kim, S.R.: Are physically embodied social agents
  better than disembodied social agents?: The effects of physical embodiment,
  tactile interaction, and people's loneliness in human--robot interaction.
  International journal of human-computer studies  \textbf{64}(10),  962--973
  (2006)

\bibitem{li2015benefit}
Li, J.: The benefit of being physically present: A survey of experimental works
  comparing copresent robots, telepresent robots and virtual agents.
  International Journal of Human-Computer Studies  \textbf{77},  23--37 (2015)

\bibitem{martini2015}
Martini, M.C., Buzzell, G.A., Wiese, E.: Agent appearance modulates mind
  attribution and social attention in human-robot interaction. In:
  International conference on social robotics. pp. 431--439. Springer (2015)

\bibitem{metta2010icub}
Metta, G., Natale, L., Nori, F., Sandini, G., Vernon, D., Fadiga, L.,
  Von~Hofsten, C., Rosander, K., Lopes, M., Santos-Victor, J., et~al.: The
  {iCub} humanoid robot: An open-systems platform for research in cognitive
  development. Neural networks  \textbf{23}(8-9),  1125--1134 (2010)

\bibitem{mollahosseini2018}
Mollahosseini, A., Abdollahi, H., Sweeny, T.D., Cole, R., Mahoor, M.H.: Role of
  embodiment and presence in human perception of robots’ facial cues.
  International Journal of Human-Computer Studies  \textbf{116},  25--39 (2018)

\bibitem{open2015estimating}
{Open Science Collaboration}: Estimating the reproducibility of psychological
  science. Science  \textbf{349}(6251),  aac4716 (2015)

\bibitem{posner1980orienting}
Posner, M.I.: Orienting of attention. Quarterly journal of experimental
  psychology  \textbf{32}(1),  3--25 (1980)

\bibitem{risko2016}
Risko, E.F., Richardson, D.C., Kingstone, A.: Breaking the fourth wall of
  cognitive science: Real-world social attention and the dual function of gaze.
  Current Directions in Psychological Science  \textbf{25}(1),  70--74 (2016)

\bibitem{roncone2016cartesian}
Roncone, A., Pattacini, U., Metta, G., Natale, L.: A {Cartesian} {6-DoF} gaze
  controller for humanoid robots. In: Robotics: science and systems. vol.~2016
  (2016)

\bibitem{stass1967eye}
Stass, J.W., Willis, F.N.: Eye contact, pupil dilation, and personal
  preference. Psychonomic science  \textbf{7}(10),  375--376 (1967)

\bibitem{tikhanoff2008open}
Tikhanoff, V., Cangelosi, A., Fitzpatrick, P., Metta, G., Natale, L., Nori, F.:
  An open-source simulator for cognitive robotics research: the prototype of
  the {iCub} humanoid robot simulator. In: Proceedings of the 8th workshop on
  performance metrics for intelligent systems. pp. 57--61 (2008)

\bibitem{wainer2006role}
Wainer, J., Feil-Seifer, D.J., Shell, D.A., Mataric, M.J.: The role of physical
  embodiment in human-robot interaction. In: ROMAN 2006-The 15th IEEE
  International Symposium on Robot and Human Interactive Communication. pp.
  117--122. IEEE (2006)

\bibitem{wiese2014using}
Wiese, E., M{\"u}ller, H.J., Wykowska, A.: Using a gaze-cueing paradigm to
  examine social cognitive mechanisms of individuals with autism observing
  robot and human faces. In: International Conference on Social Robotics. pp.
  370--379. Springer (2014)

\bibitem{wiese2018}
Wiese, E., Weis, P.P., Lofaro, D.M.: Embodied social robots trigger gaze
  following in real-time hri. 2018 15th International Conference on Ubiquitous
  Robots (UR) pp. 477--482 (2018). \doi{10.1109/URAI.2018.8441825}

\bibitem{wiese2012see}
Wiese, E., Wykowska, A., Zwickel, J., Müller, H.J.: I see what you mean: How
  attentional selection is shaped by ascribing intentions to others. PLOS ONE
  \textbf{7}, ~1--7 (09 2012). \doi{10.1371/journal.pone.0045391},
  \url{https://doi.org/10.1371/journal.pone.0045391}

\bibitem{wykowska2014beliefs}
Wykowska, A., Wiese, E., Prosser, A., M{\"u}ller, H.J.: Beliefs about the minds
  of others influence how we process sensory information. PloS one
  \textbf{9}(4),  e94339 (2014)

\end{thebibliography}

\end{document}